\documentclass[aps,twocolumn,pra,superscriptaddress,notitlepage,floatfix]{revtex4-2}
\pdfoutput=1

\usepackage{graphicx,color,xcolor}
\usepackage{marvosym}
\usepackage{amsmath,amssymb}
\usepackage{ulem}

\usepackage{graphicx,amsmath,graphics}
\usepackage{amssymb,exscale}
\usepackage{mmap} 
\usepackage[colorinlistoftodos]{todonotes}
\usepackage[colorlinks=true, allcolors=blue]{hyperref}

\newcommand{\ie}{i.\,e.}%

\makeatletter
\def\subsubsection{\@startsection{subsubsection}{3}{10pt}{-1.25ex plus -1ex minus -.1ex}{0ex plus 0ex}{\normalsize\bf}}
\def\paragraph{\@startsection{paragraph}{4}{10pt}{-1.25ex plus -1ex minus -.1ex}{0ex plus 0ex}{\normalsize\textit}}
\renewcommand\@biblabel[1]{#1}
\renewcommand\@makefntext[1]%
{\noindent\makebox[0pt][r]{\@thefnmark\,}#1}
\DeclareRobustCommand\onlinecite{\@onlinecite}
\def\@onlinecite#1{\begingroup\let\@cite\NAT@citenum\citealp{#1}\endgroup}
\def\tagform@#1{\maketag@@@{\ignorespaces#1\unskip\@@italiccorr}}
\let\orgtheequation\theequation
\def\theequation{(\orgtheequation)}
\makeatother

\newcommand{\expectation}[3]{\left\langle #1\left|#2\right|#3\right\rangle}
\newcommand{\expected}[1]{\left\langle #1\right\rangle_t}

\newcommand{\blue}[1]{{\color{blue}#1}}

\newcommand{\granadaD}{Departamento de F\'{i}sica At\'{o}mica, Molecular y Nuclear, Universidad de Granada, 18071 Granada, Spain}
\newcommand{\granadaI}{Instituto Carlos I de F\'{i}sica Te\'{o}rica y Computacional, Universidad de Granada, 18071 Granada, Spain}
\newcommand{\hhZ}{Zentrum f\"ur Optische Quantentechnologien, Universit\"at Hamburg, Luruper Chaussee 149, 22761 Hamburg, Germany}
\newcommand{\hhCUI}{The Hamburg Centre for Ultrafast Imaging, Universit\"at Hamburg, Luruper Chaussee 149, 22761 Hamburg, Germany}

\begin{document}

\title{Flipping electric dipole in the vibrational wave packet dynamics of carbon monoxide}

\author{Carlos Barbero-Petrel}
\affiliation{\granadaD}

\author{Peter Schmelcher}
\affiliation{\hhZ}
\affiliation{\hhCUI}

\author{Rosario~Gonz\'{a}lez-F\'{e}rez}
\affiliation{\granadaD}
\affiliation{\granadaI}

\date{\today}

\begin{abstract}
Recently Rydberg atom-ion bound states have been observed using a high resolution ion microscope 
(Nature {\bf{605}}, 453 (2022)) and the corresponding vibrational dynamics has been spectroscopically analyzed.
The atom-ion bond is created by an avoided crossing, which involves a flipping molecular dipole.
Motivated by the discovery of this binding mechanism we address here the question whether a similar
behavior can also occur for ground state diatomic molecules.
Specifically, we investigate the vibrational wave packet dynamics within the 
$^1\Sigma^+_g$ electronic ground-state of carbon monoxide (CO), 
which shows a zero crossing of its dipole moment function close to its equilibrium. 
Via time-evolution of coherent states we demonstrate that indeed a flipping dipole is obtained and
its dynamics can be controlled to some extent. Varying the coherent state parameter we explore different regions
of the vibrational excitation spectrum thereby tuning the time scales of the rapid oscillatory
motion of the relevant observables, their decay and revivals as well as the transition
to a regime of irregular dynamics.
\end{abstract}


\maketitle

\section{Introduction}
\label{sec:intro}

As a consequence of the many decade-long investigations on the theory of the electronic structure of
molecules~\cite{Helgaker13,Szabo96} we have now a profound understanding of the molecular binding mechanisms
including the paradigms of covalent, ionic and correlation-based binding. While these achievements have
enabled us to steadily advance to increasingly larger molecular systems thereby covering more and more of
the chemically relevant landscape of molecules by using in particular coupled cluster~\cite{Bartlett07}
and density functional theory~\cite{Mardirossian17} a parallel development has taken place, which opens up
the route of exploring synthetic quantum systems. One major category of synthetic quantum systems are those
based on neutral ultracold quantum matter~\cite{Pethick11}. Cooling, trapping and controlling the interactions
among atoms~\cite{Grimm00,Chin10} allows not only to prepare designated many-body quantum systems~\cite{Lewenstein07}, 
which is at the heart of quantum simulation~\cite{Lewenstein12}, but opens also the pathway of 
preparing novel kinds of few-body atomic and molecular systems. Among these are weakly bound 
ultracold species, such as Efimov states~\cite{Greene17} but in particular also molecular species
that have highly-excited Rydberg atoms as a basic ingredient~\cite{Gallagher94}. One outstanding
category of species are the so-called ultralong-range Rydberg molecules composed of a Rydberg atom
and one to many neutral ground state atoms~\cite{Fey20}. These highly asymmetric and hugely extended
molecules were first predicted theoretically~\cite{Greene00} and subsequently observed experimentally almost a 
decade later~\cite{Bendkowsky09}. They can be divided into low angular momentum non-polar and high angular
momentum polar Rydberg molecules where the binding is established through the low-energy scattering
of the Rydberg electron off the ground state atoms. A plethora of intriguing properties of these 
ultralong-range molecules have been found in the past decade rendering them a laboratory for
novel binding mechanisms and dynamical processes on extraordinary length and time 
scales~\cite{Bendkowsky10,Li11,Tallant12,Balewski13,Gaj14,DeSalvo15,Camargo18,Engel19,Hummel20}.

Very recently the binding of an atomic ion to a Rydberg atom has been predicted \cite{Duspayev21,Deiss21} and
experimentally observed~\cite{Zuber22}. The underlying binding mechanism is different from
the above-mentioned ones. The positively charged ion interacts with the Rydberg dipole moment, 
which varies substantially across the potential well that contains the corresponding 
vibrational bound states. The Rydberg atoms dipole moment thereby covers a range of negative and positive values
due to the fact that an avoided crossing between low- and high-angular momentum states is
responsible for the formation of the underlying potential well. This has been resolved
in an ion microscope thereby imaging the radial extent and angular alignment of the molecule~\cite{Zuber22}.
In a next step the vibrational and orientational dynamics of the Rydberg atom-ion-molecules has been 
probed~\cite{Zou23} using the high resolution ion microscope. 

Motivated by the above-mentioned analysis of the bonding mechanism of the Rydberg atom-ion-molecule the natural
question appears whether similar properties could also appear for 'traditional' ground state diatomic
molecules. More specifically, could one obtain an oscillating and flipping dipole for a vibrational wave
packet oscillating in the corresponding ground state potential energy curve shown in~\autoref{fig:V_D_R}~(a). To answer this question
we address in this work the wave packet dynamics taking place in the potential energy curve of the
electronic $^{1}\Sigma_{g}^{+}$ ground state of the carbon monoxide molecule (CO). The latter is known to change the
character of its bond from covalent around the equilibrium position to ionic for larger internuclear distances.
More specifically, we choose CO as a prototype due to its electric dipole moment (EDM) function
$D(R)$ shown in~\autoref{fig:V_D_R}~(b). The EDM possesses a zero crossing close to its
equilibrium internuclear distance (see~\autoref{fig:V_D_R}~(a)) and becomes substantially negative and positive for smaller
and larger internuclear distances, respectively. It reaches a maximum for an internuclear distance
which is almost two times the equilibrium one. As a consequence, it has been shown in a previous
work that the impact of an electric field strongly depends on its vibrational excitation:
the energetically lowest-lying states become anti-oriented with respect to the field direction,
whereas higher excited ones are oriented~\cite{Gonzalez04}.

In the present work we employ initial wave packets that are coherent states of a harmonic oscillator potential 
related to the potential energy curve of the electronic ground state of CO and propagate them in time
to explore their dynamical properties. Depending on the parameter values of the coherent states they cover
different regimes of the vibrational eigenspectrum and possess an oscillating EDM that changes sign in
the course of its dynamical evolution. Our analysis covers the complete dynamics including 
short time oscillations, decay and (fractional) revivals accompanied by a characteristic spreading and
refocusing of the wave packet as well as, and in particular, their characteristic dipole dynamics.

The paper is organized as follows. In~\autoref{sec:Hamil}, we describe the Hamiltonian, initial
wave packet and the expectation values used to explore the corresponding time evolution.
In~\autoref{sec:vib_dynamics}, the vibrational wave packet dynamics is analyzed with varying coherent
state parameters thereby addressing regimes of different dynamical behaviour. Our conclusions are
presented in~\autoref{sec:con}.

\section{Vibrational dynamics and coherent state}
\label{sec:Hamil}

\begin{figure}[t]
\centering
\includegraphics[width=0.98\columnwidth]{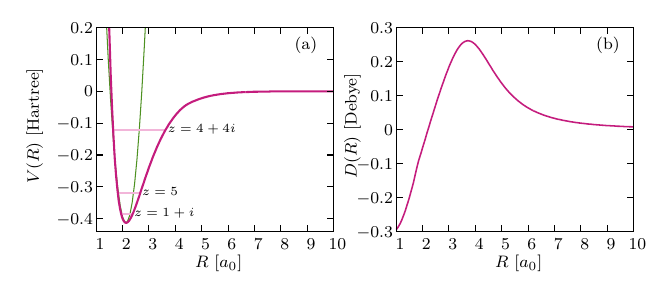}
\caption{For carbon monoxide, we show (a) the electronic ground-state potential curve (thick pink line),
and the corresponding fitted harmonic oscillator potential (thin green line), and (b) the electric dipole moment
function. In panel (a) the horizontal lines indicate the energies of the coherent states together with their parameter
values $z$.}
\label{fig:V_D_R}
\end{figure}

Within the Born-Oppenheimer approximation, the Hamiltonian  of the rovibrational  motion of a diatomic molecule reads
\begin{equation}
\label{Hamilt}
H = -\frac{\hbar^2}{2\mu R^2} \frac{\partial}{\partial R} \left(R^2\frac{\partial}{\partial R} \right) + \frac{\mathbf{N}^2}{2\mu R^2} +
 V(R) 
\end{equation}
where the first and second terms represent the vibrational and rotational kinetic energies, 
$\mathbf{N}$ is the rotational angular momentum, $R$ the internuclear coordinate,
and $\mu$  the reduced mass. 
The electronic potential energy curve $V(R)$ of the ground-state of CO is presented in~\autoref{fig:V_D_R}~(a) \cite{LeFloch}.
Here, we focus on states with no rotational excitations $N=0$, and analyze the corresponding
vibrational wave packet dynamics with an emphasis on the dynamics of the dipole of the wave packet.

In addition to the reduced (w.r.t. the integration measure involving $R$)
Schr\"odinger equation belonging to the Hamiltonian~\ref{Hamilt} and the 
resulting vibrational eigenstates, we employ complementary a one-dimensional harmonic oscillator potential 
that approximates the electronic ground-state potential of CO around its equilibrium.
Our initial wave packet is a coherent state of this fitted harmonic oscillator potential.
In position representation this coherent state reads~\cite{Galindo89,Scully}
\begin{eqnarray}
\nonumber
\psi_z(R) &=& \sqrt{\frac{\alpha}{\sqrt{\pi}}} \exp \left[-\frac{(|z|^2-z^2)}{2}\right] \\
&\times& \exp \left[-\frac{(\alpha (R-R_{0})-\sqrt{2}z)^2}{2} \right],
\label{coherent_wave}
\end{eqnarray}
where  $\alpha = \sqrt{\mu\omega/\hbar}$ is the inverse of the harmonic oscillator length,
$\omega$ the frequency of the fitted harmonic oscillator, and  $R_{0}=2.13~a_0$ corresponds to the position of the
minimum of the electronic ground state potential energy curve of CO.
The coherent state  is an eigenstate of the annihilation operator, and $z$, which can take
complex values, is the corresponding eigenvalue. In this work, we explore the wave packet dynamics
of this coherent state by varying $z$ and thereby cover different dynamical regimes.

For a given value of $z$, the  coherent state is expanded in the, admittedly incomplete, basis formed by the
vibrational eigenfunctions $\varphi_\nu(R)$ of the CO molecule as 
\begin{equation}
\label{coherent_decomp}
\psi_z (R) = \sum_{\nu=0}^{\nu_{max}} c^z_{\nu} \varphi_{\nu}(R), 
\end{equation}
where $\nu$ is the vibrational quantum number and $c^z_{\nu} =\langle  \varphi_{\nu}(R)|  \psi_z (R) \rangle$. 
The vibrational (reduced) eigenfunctions $\varphi_\nu(R)$
are obtained by solving the time-independent Schrödinger equation belonging to the Hamiltonian in \autoref{Hamilt}.
Note that the electronic ground state of  CO possesses more than $80$ bound vibrational levels for $N=0$, and in
our calculations we take $\nu_{max}=82$. 
However, only the set of vibrational states with  $|c^z_{\nu}|\gtrsim 0.01$ have a significant impact on the dynamics, 
and this set strongly depends on the chosen value of $z$. 
Of course, the coherent state is only approximated
by the expansion~\ref{coherent_decomp} and, however, we cover the coherent state with an accuracy
above $99.99\%$ of the corresponding overlap.

Since the Hamiltonian is time-independent, the time evolution of the coherent state is given by
\begin{equation}
\label{temp_evo}
\Psi_z (R,t) = \sum_{\nu=0}^{\nu_{max}}c^z_{\nu} \varphi_{\nu}(R) e^{-itE_{\nu}/\hbar} \, ,
\end{equation}
where $E_\nu$ is the eigenvalue associated to the eigenstate $\varphi_\nu(R)$.  
The vibrational dynamics is analyzed by inspecting the expectation values 
$\expected{R}=  \expectation{\Psi_z(R,t)}{R}{\Psi_z(R,t)}$ and $\expected{D(R)}=  \expectation{\Psi_z(R,t)}{D(R)}{\Psi_z(R,t)}$,
with $D(R)$ being the EDM of CO~\cite{Chackerian}. Their temporal evolution is encapsulated in the general equation
\begin{eqnarray}
&&\expected{f(R)}= \expectation{\Psi_z(R,t)}{f(R)}{\Psi_z(R,t)} =  
\\
&&\sum_{\nu=0}^{\nu_{max}}|c_{\nu}|^2\langle f(R)\rangle_{\nu\nu}+2\sum_{\nu\ne\nu^\prime}c_{\nu^\prime}^*c_{\nu}\cos\left(\omega_{\nu^\prime\nu} t\right) \nonumber
\langle f(R)\rangle_{\nu^\prime\nu}
\end{eqnarray}
with  $f(R)=R$, $D(R)$ in our case, the vibrational frequency being 
$\omega_{\nu^\prime\nu}=(E_{\nu^\prime}-E_{\nu})/\hbar$, and
\begin{equation}
\label{exp_vibra}
\langle f(R)\rangle_{\nu'\nu}= \int \varphi_{\nu^\prime}^*(R) f(R)\varphi_{\nu}(R) dR,
\end{equation}
Thus, $ \expected{f(R)}$  oscillates around the  average of the contributions of the populated vibrational
levels, $\langle f(R)\rangle_{\nu\nu}$,  weighted by $|c_{\nu}|^2$.  
The oscillatory behavior is governed by cross terms with frequencies $\omega_{\nu^\prime\nu}$. 
In contrast to the constant rotational spacing of a rigid molecule, which determines also the rotational dynamics 
in a constant electric or laser field, the vibrational spacing typically decreases with increasing
degree of excitation and the corresponding frequencies $\omega_{\nu^\prime\nu}$ do not form a commensurate set.
We also analyze the temporal evolution of the spatial delocalization of the wave packet using the variance 
\begin{equation}
\label{fig:sigma}
\sigma_R(t)= \sqrt{\expected{R^2}- \expected{R}^2} \,.
\end{equation}

\section{Vibrational wave packet dynamics}
\label{sec:vib_dynamics}

Our fit of a harmonic oscillator potential accounts properly for the repulsive barrier of the potential energy curve
of CO,  \ie, for $R<R_{min}$, see~\autoref{fig:V_D_R}~(a). For its frequency we obtain 
\blue{$89\pm16$~THz}.
This ensures that the spreading of the wave packet remains within the CO potential energy curve.
  
In the coherent state~\autoref{coherent_wave}, 
the two degrees of freedom of the parameter $z$, \ie, its real and imaginary parts $z_{R}$ and $z_{I}$,  play a similar 
role, and qualitatively similar results are obtained by keeping  $|z|$ constant.
By increasing $|z|$, the center of the initial coherent state is shifted towards larger values of $R$, which implies that 
higher energies within the vibrational spectrum are explored by the corresponding wave packet.
In the next sections, we analyze the vibrational dynamics for the coherent states with  $z=1+i$, $z=5$ and $z=4+4i$. 

\begin{figure}[t]
\includegraphics[width=0.95\columnwidth]{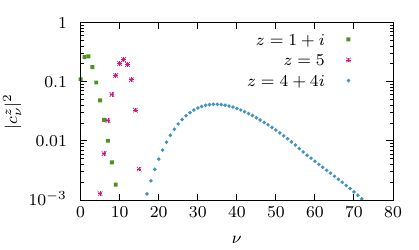}
\caption{Absolute value squared of the coefficients  $|c_\nu^z|^2$ of the 
decomposition~\autoref{coherent_decomp} in terms of the vibrational eigenstates of CO, 
for the coherent states with $z=1+i$ (green squares) $z=5$ (pink stars) and $z=4+4i$  (blue circles).}
\label{fig:weights}
\end{figure}

\subsection{Wave packet dynamics for $z=1+i$} \label{z_1_i_case}

\begin{figure}[t]
\centering
\includegraphics[width=0.95\columnwidth]{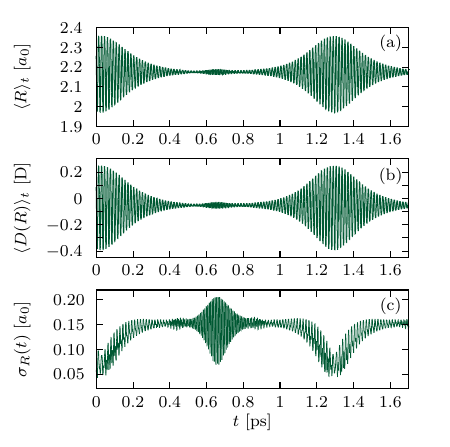}
\caption{Wave packet dynamics of the $z=1+i$ coherent state. Shown is the time evolution 
of the expectation values (a) $\expected{R}$ and (b) $\expected{D(R)}$, as well as (c) the variance $\sigma_R(t)$.}
\label{fig_z1}
\end{figure}

For the $z=1+i$ coherent state, only the lowest ten vibrational states of CO 
possess a weight larger than $10^{-3}$ in the expansion~\eqref{temp_evo}, 
with the vibrational eigenstates up to the third excited one having the dominant
contributions as shown in~\autoref{fig:weights}.  
The energy of this wave packet is relatively small, $-0.386$~Hartree, and 
lies between the  $\nu=3$ and $\nu=4$ eigenstates of CO.
Hence, its vibrational dynamics is confined to the lower part of the CO potential energy curve as illustrated 
in~\autoref{fig:V_D_R}~(a).
The time evolutions of  $\expected{R}$, $\expected{D(R)}$, and $\sigma_R(t)$  are presented in~\autoref{fig_z1}.
$\expected{R}$ and  $\expected{D(R)}$ show an oscillatory behavior around the mean values $2.18~a_0$ and 
$-0.05$~D, respectively. 
Most important, the EDM  of this wave packet oscillates rapidly between
positive and negative values, the maximum value being $\expected{D(R)}=0.24$~D  and the minimum 
one is $\expected{D(R)}=-0.39$~D.

\begin{figure}[t]
\centering
\includegraphics[width=0.98\linewidth,angle=0]{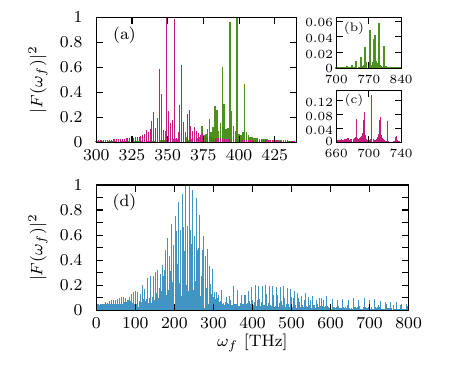}  
\caption{Power spectrum obtained from the discrete Fourier transform of $\expected{R}$ for the coherent states 
(a) $z=1+i$ (thick green line) and $z=5$ (thin pink line), and for (d) $z=4+4i$ (thin blue line).
Panels (b) and (c) show the power spectrum for a window of larger frequencies for $z=1+i$ and $z=5$ wave packets, respectively.  
Note that $|F(\omega_f)|^2$  is normalized such that the maximum value equals one.}
\label{fig:fourier_z_1_1}
\end{figure}

The similar behavior of $\expected{R}$ and  $\expected{D(R)}$ is due to the almost linear dependence
of $D(R)$ on $R$ in the relevant range of $V(R)$ covered by the vibrational dynamics (see~\autoref{fig:V_D_R}~(b)).
They both show rapid oscillations, with a frequency of approximately $393.5$~THz, and a period of approximately $0.016$~ps.
The amplitudes are
modulated by a slower frequency corresponding to $4.9$~THz, which is responsible for the long time revival
at $t\approx 1.28$~ps. The discrete Fourier transform of $\expected{R}$ is presented in 
Figs.~\ref{fig:fourier_z_1_1}~(a) and~(b).
The frequencies due to nearest neighbor vibrational states $\omega_{\nu+1,\nu}$ are dominant and can
clearly be observed in Fig.~\ref{fig:fourier_z_1_1}~(a).
Indeed, the large and small frequencies dominating the vibrational dynamics 
are approximately given by $(\omega_{1,2}+\omega_{2,3})/2$ and 
$\omega_{2,3}-\omega_{1,2}$, with $\omega_{1,2}=398.7$~THz and $\omega_{2,3}=393.8$~THz.
The frequencies due to the next nearest vibrational neighbors, $\omega_{\nu+2,\nu}$, are shown in panel~(b)
of~\autoref{fig:fourier_z_1_1}, and their influence on the dynamics is rather small.

\begin{figure}[t]
\includegraphics[width=0.98\linewidth,angle=0]{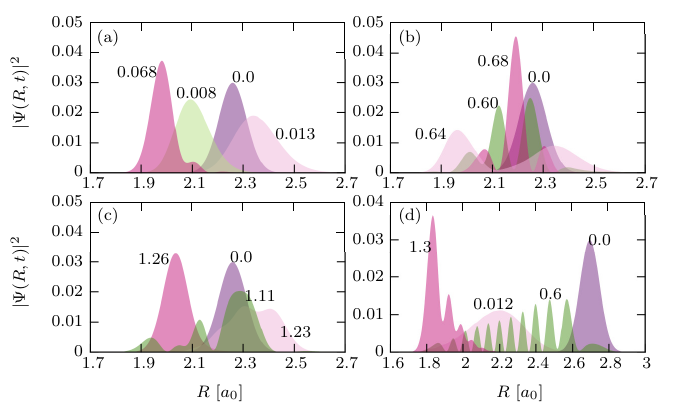}
\caption{The absolute value squared of the wave packet at different times, indicated in units of picoseconds in the panels, 
(a), (b) and (c) for the  $z=1+i$ coherent state, and (d) for the $z=5$ coherent state.}
\label{1_cycle_1}
\end{figure}

The spreading of the wave packet is illustrated in~\autoref{fig_z1}~(c)
via the variance $\sigma_R(t)$ defined in~\autoref{fig:sigma}.
This is complemented by snapshots of the square of the absolute value 
of the wave packet for several time instants in~\autoref{1_cycle_1}. 
During the first oscillations, the wave packet maintains its Gaussian shape, 
but increases its FWHM and decreases its height while its center oscillates around the
minimum of the potential well, as it is illustrated in panel (a) of~\autoref{1_cycle_1} for the time
instants $t= 0.0, 0.008$ and  $0.013$~ps. 
For  $t=0.068$~ps, the wave packet has become narrower, whereas its 
Gaussian shape starts to be more strongly distorted, manifesting itself in the existence 
of a pronounced asymmetry and a right-sided tail with a second maximum being visible. 

Following the decay of the amplitude of the fast oscillations of $\expected{R}$ and $\expected{D(R)}$
(see~\autoref{fig_z1}~(a,b)) we observe for the time window $0.5$~ps $\lesssim t \lesssim0.9$~ps
via the large values and large amplitude oscillations of the variance in~\autoref{fig_z1}~(c)
a substantial spreading (and partial refocusing) of the wave packet. This is confirmed
by the snapshots of the absolute square of the wave packet presented in~\autoref{1_cycle_1}~(b). 
It is now spread over the entire region between the classical turning points of the potential, and 
exhibits several maxima, due to the contribution of several vibrational eigenstates.
With further increasing propagation time, the mean value of $\sigma_R(t)$  first remains approximately
constant but then starts to decrease for $t\gtrsim 1.1$~ps thereby approaching the first revival. 
In this region, the wave packet is initially spread over large parts of the potential
but becomes more localized again during the revival as illustrated in~\autoref{1_cycle_1}~(c) for $t=1.26$~ps.

\subsection{Wave packet evolution for $z=5$}
\label{z_5_0_case}

\begin{figure}[t]
\includegraphics[width=0.95\columnwidth]{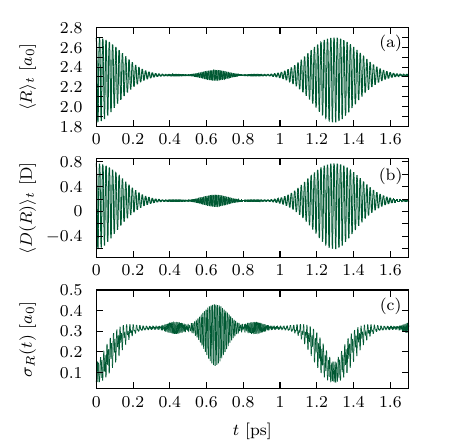}
\caption{
Wave packet dynamics of the $z=5$ coherent state. Time evolution 
of the expectation values (a) $\expected{R}$ and (b) $\expected{D(R)}$, as well as (c) the variance 
$\sigma_R(t)$.}
\label{fig_z5}
\end{figure}

For the $z=5$ coherent state, higher excited eigenstates contribute substantially to its time evolution~\eqref{temp_evo},
and the dominant vibrational excitations are those for $10\le \nu \le 13$, see~\autoref{fig:weights}.
The wave packet possesses a larger energy $-0.319$~Hartree, \ie, it lies energetically 
between the $\nu=9$ and $\nu=10$ vibrational eigenstates. Thus, a larger range within the potential energy
curve is accessible for the vibrational dynamics as illustrated in~\autoref{fig:V_D_R}~(a). 
The dynamics of the observables $\expected{R},\expected{D(R)}$
in~\autoref{fig_z5} shows a qualitatively similar but quantitatively different behavior as for the
$z=1+i$ coherent state in Fig.\ref{fig_z1}.
The expectation values $\expected{R}$ and $\expected{D(R)}$ rapidly oscillate in time
with a smaller frequency of $351$~THz, and these oscillations possess larger amplitudes due to the contribution 
of higher excited vibrational states.
These fast oscillations are again modulated by a smaller frequency of $4.9$~THz, contributing to the observed
decay and revival. The averages of $\expected{R}$ and $\expected{D(R)}$
are shifted to  larger values  $2.31~a_0$ and $ 0.17$~D, respectively. 
As for the $z=1+i$ coherent state,  $\expected{D(R)}$ shows also for $z=5$ oscillations corresponding
to an alternating sign of the dipole moment with a maximum value of $0.77$~D and a minimum value of $-0.60$~D. 

The frequencies obtained from the discrete Fourier transform of $\expected{R}$ are presented in 
Figs.~\ref{fig:fourier_z_1_1}~(a) and~(c). 
The energy gap between neighboring vibrational levels is reduced as we move energetically up in the CO vibrational spectrum.
This explains that the dominant frequencies in $\expected{R}$ are smaller compared to those of the $z=1+i$ coherent state, 
see the pink versus green histograms in~\autoref{fig:fourier_z_1_1}~(a).
The rapid oscillations of  $\expected{R}$ can be explained in terms of $(\omega_{11,12}+\omega_{12,13})/2$ and 
the frequency modulating their amplitude is $\omega_{12,13}-\omega_{11,12}$, with $\omega_{11,12}=354.4$~THz and 
$\omega_{12,13}=349.6$~THz.

The variance in~\autoref{fig_z5}~(c) indicates a larger spreading of the wave packet, which is also illustrated by
the snapshots at several times in~\autoref{1_cycle_1}~(d). 
The spatial width of the wave packet is larger, and the contribution of higher excitations is reflected 
in the presence of more maxima as a function of the internuclear distance $R$. 

\subsection{Wave packet dynamics for $z=4+4i$.}

By further increasing  $|z|$, the decomposition~\eqref{coherent_decomp} of the coherent state involves
more eigenstates with a significant weight. This implies that more frequencies participate in the
resulting dynamics and, as a consequence, the characteristic oscillations of the expectation values observed
in the previous cases might disappear. 
For instance, in the expansion~\eqref{coherent_decomp} of the $z=4+4i$ wave packet the eigenstates
within the vibrational range $28\le\nu\le 55$ ($18\le\nu\le72$) possess weights larger than $10^{-2}$ ($10^{-3}$),
see~\autoref{fig:weights}.  
Energetically, this  coherent state with an energy $-0.121$~Hartree lies between the $\nu=37$ and $\nu=38$ 
vibrational eigenstates, and has a larger internuclear region to explore as shown in~\autoref{fig:V_D_R}~(a).

\begin{figure}[t]
\includegraphics[width=0.95\columnwidth]{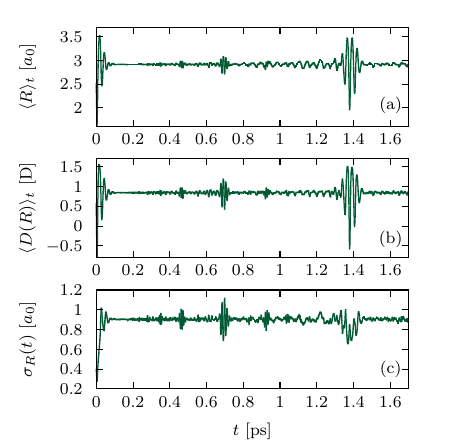}
\caption{Wave packet dynamics of the $z=4+4i$ coherent state. Time evolution 
of the expectation values (a) $\expected{R}$ and (b) $\expected{D(R)}$, as well as (c) the variance 
$\sigma_R(t)$.}
\label{fig_z4}
\end{figure}

This large number of vibrational states with similar weights significantly impacts the dynamics as illustrated by 
the time evolutions of $\expected{R}$ and $\expected{D(R)}$ 
in~\autoref{fig_z4}~(a) and (b), respectively. $\expected{R}$ and $\expected{D(R)}$ 
do not show regular oscillations due to the many frequencies with a similar contribution. 
A rapid decay within the first few oscillations is observed
followed by an irregular behavior with occasionally occurring partial revivals.
The corresponding power spectrum of $\expected{R}$ is presented in~\autoref{fig:fourier_z_1_1}~(d)
demonstrating the broad bands of contributing frequencies.
Severe differences occur in comparison to the power spectrum of the wave packets belonging
to $z=1+i$ and $z=5$ presented in panel (a) can be observed. For example, and among others,
there are now six frequencies with a contribution larger than 0.8.

A few more specific remarks are in order.
Initially $\expected{R}$ and $\expected{D(R)}$ show rapid oscillations, in which $\expected{D(R)}$ is, during
a certain interval of the oscillatory motion, negative for approximately only $6$~fs.
After these few fast oscillations at short times, the dipole moment remains approximately constant and positive
with small amplitude deviations. As expected, due to the participation of higher excited vibrational eigenstates
in the wave packet dynamics, the mean values of $\expected{R}$ and $\expected{D(R)}$ are larger as compared
to the previously analyzed coherent state dynamics.
The observed half and full revivals remain as characteristic features of the dynamics, and between them several irregular 
oscillatory bursts of motion are encountered. During the full revival, $\expected{D(R)}$ is negative for the
very short period of $5$~fs only. These short time intervals where $\expected{D(R)}$ changes sign and
becomes negative also appear for coherent states with even larger values of $|z|$. 
The variance $\sigma_R(t)$, shown  in~\autoref{fig_z4}~(c), also presents a distinctly different time evolution compared to the
previous cases. Its large value indicates that the wave packet is delocalized and spreads over large parts of the 
potential energy curve in the course of the dynamics.  

\section{Conclusions}
\label{sec:con}

Motivated by the recently found Rydberg atom-ion molecules with a molecular bond that shows a
flipping dipole~\cite{Zuber22,Zou23} we explore in this work the vibrational dynamics of a wave
packet within the electronic ground state of the CO molecule. Our specific choice is well-grounded in the
fact that CO possesses an electronic dipole moment function that has (approximately) a zero crossing
at the equilibrium internuclear distance, with negative/positive dipole moments for smaller/larger internuclear
separations. We have constructed coherent wave packets for different parameter values as initial states for
our study of the dynamical evolution. Several relevant observables such as the mean internuclear separation
$\expected{R}$, the dipole moment $\expected{D(R)}$
as well as the variance $\sigma_R(t)$ of the wave packet have been analyzed.

For low and moderate values of the coherent state parameter $|z|$,
the coherent state lies in the lower part of the vibrational spectrum of CO.
It can be expressed as a superposition of a relatively small number of vibrational states, and the 
time evolutions of  $\expected{R}$ and $\expected{D(R)}$ are dominated by two frequencies. 
Consequently the wave packet performs spatial oscillations in time close to the potential minimum where $D(R)$ changes
its sign. Therefore, we encounter a flipping dipole moment taking on both positive and negative values during
the corresponding phases of the evolution. Such a behavior cannot be realized on basis of a
pure rotational dynamics including electric or non-resonant laser fields due to the smaller energy scales involved.
Apart from fast oscillations we encounter a slow decay and subsequent (fractional) revivals of the
wave packet which is accompanied by a similar characteristic behavior of the dipole moment.

By increasing $|z|$, the wave packet possesses a larger energy, explores a larger internuclear range, and 
higher and more vibrational excitations control the dynamics.
The many contributing non-commensurable vibrational frequencies lead to a rapid decay of the fast
oscillations of $\expected{R}$ and $\expected{D(R)}$, but still the half and full revivals
appear as characteristic features of their time evolution, although taking place at irregular time intervals.
Interestingly, $\expected{D(R)}$, which is now positive, reaches negative values only during the case of
a complete revival.

A few comments concerning the experimental implementation of our
setup are in order. The preparation of the coherent state of the
CO vibrational dynamics could be accomplished by ensuring the proper
populations of the (few most relevant) excited states, ideally using an optimally designed
light pulse providing the proper phase relationship~\cite{Pierce,Werschnik,Brif}. Concerning the
observation of the coherent state dynamics and in particular the
flipping dipole dynamics a pump probe spectroscopy might pave the
way as it is used frequently in the literature~\cite{Parson}. 

Taking the present work as a starting-point for future investigations, one could envisage to explore
other diatomic systems with a similar or complementary structure of the dipole moment function. It is
an open route of exploration to construct corresponding wave packets for polyatomic systems with several
vibrational degrees of freedom that might show an even more intriguing dynamics of the corresponding
dipole or higher moments.

\begin{acknowledgments} 
C.B. and R.G.F. gratefully acknowledge financial support by the Spanish projects PID2020-113390GB-I00 (MICIN), and 
PY20$\_$00082 (Junta de Andalucía),   and the Andalusian research group FQM-207.
\end{acknowledgments}

\end{document}